# Estimation of the Adriatic sea water turnover time using fallout $^{90}$Sr as a radioactive tracer

Zdenko FRANIC

Institute for Medical Research and Occupational Health, Radiation Protection Unit, HR-10000 Zagreb, Ksaverska cesta 2, PO Box 291, Croatia

E-mail: franic@franic.info

**Abstract**

Systematic, long term measurements, starting in 1963, of $^{90}$Sr activity concentrations in sea water have been performed at four locations (cities of Rovinj, Rijeka, Split and Dubrovnik) along the Croatian coast of the Adriatic sea. In addition, fallout samples were collected in the city of Zadar. $^{90}$Sr activity concentrations are in good correlation with the fallout activity, the coefficient of correlation being 0.72. After the nuclear moratorium on atmospheric nuclear bomb tests in 1960s, $^{90}$Sr activity concentrations in sea water exponentially dropped from $14.8 \pm 2.4$ Bq m$^{-3}$ in 1963 to $2.0 \pm 0.3$ Bq m$^{-3}$ in 2003. In the same period, the total annual $^{90}$Sr land surface deposit in Zadar fell by three orders of magnitude, from 713.3 Bq m$^{-2}$ in 1963 to 0.4 Bq m$^{-2}$ in 2003. Using strontium sea water and fallout data, a mathematical model was developed to describe the rate of change of $^{90}$Sr activity concentrations in the Adriatic sea water and estimate its mean residence time in the Adriatic. By fitting the experimental data to a theoretically predicted curve, the mean residence time of $^{90}$Sr in the Adriatic sea water was estimated to be approximately $3.4 \pm 0.4$ years, standard deviation being calculated by Monte Carlo simulations. As in physical oceanography $^{90}$Sr can be used as effective radioactive tracer of water mass transport, this value also reflects the upper limit for turnover time of the Adriatic sea water. The turnover time of 3.4 years for the Adriatic sea water is in reasonable agreement with the value which was estimated, by studying water flows through the Strait of Otranto, to be on the order of 1 year.

# 1. Introduction

## 1.1. Position, topography and characteristics of the Adriatic sea

The Adriatic sea, extending to 45° 47' N, is the northernmost part of the Mediterranean sea, belonging to the eastern Mediterranean basin. It is located between Italy and the Balkan Peninsula. As shown on Fig. 1, it is landlocked on the north, east and west, and is linked with the Mediterranean through the Strait of Otranto to the south.

**Figure 1 about here**

The Adriatic is a rectangularly shaped basin, oriented in an NW-SE direction with a length of about 800 km and a width varying from 102 to 355 km, the average being about 250 km. Northern half of the Adriatic can be divided into two sub-basins:

a) A northernmost shallow basin with the bottom sloping gently to the south and reaching at most 100 m, then dropping quickly to 200 m just south of Ancona and

b) Three pits located along the transversal line of the Italian city of Pescara, one of which is known as the Jabuka Pit.

Southern half of the Adriatic consists of a basin, called the South Adriatic Pit which is separated from the middle basin by the 170 m deep Palagruza Sill. It is characterized by approximately circular isobaths, with a maximum depth of about 1200 m in the center.

The bottom rises toward the Strait of Otranto past the southern basin, with the strait having a maximum depth of 780 m, and average depth of 325 m, and a width of about 75 km. Bottom depth of the Adriatic sea is shown on Fig. 2.

**Figure 2 about here**

The western coast of the Adriatic sea is regular, with isobaths running parallel to the shoreline and depth increasing uniformly seawards. The more rugged eastern coast is composed of many islands and headlands rising abruptly from the deep coastal water.

Physical characteristics of the Adriatic sea (Leksikografski institut, 1979) are summarized in Table 1.

**Table 1 About here**

*1.2. Hydrology and meteorology*

Geomorphological characteristics of the Adriatic sea (elongated shape, almost land locked position between the mountains situated on Balkan and Italian Peninsulas and relatively shallow waters, especially in northern part) play important part in controlling the dynamics of its waters. Due to its landlocked position, the Adriatic sea is subject to highly variable atmospheric forcing. As a result, the oceanographic properties of the Adriatic sea, like circulation and distribution of its water masses strongly depend on the characteristics of air-sea fluxes (Cushman-Roisin, 2001).

The long-term average runoff rate along the Adriatic coast is 5500 - 5700 $m^3 s^{-1}$, the Po river carrying alone 28% of the total runoff, i.e., 1540 - 1600 $m^3 s^{-1}$. (Cushman-Roisin et al., 2001; Sekulic and Vertacnik, 1996; Raicich 1996.) It can easily be calculated that on annual scale, total runoff corresponds to annual addition of about 1.3 m thick water layer over the whole basin, which is approximately 0.5 % of the total Adriatic sea volume.

The water exchange pattern between the Adriatic and the greater Mediterranean through the Strait of Otranto, suggests an inflow along the eastern and outflow along the western coast. The outflowing portion of the water exchange though the strait area, consists of a surface Adriatic water and cold outflowing vein of the dense water formed in the Adriatic. The inflowing water from the Mediterranean origin is more saline and warmer. Surface currents are responsible for the transports of an important part of the marine pollutants and for the freshwater dispersion. The circulation regime varies seasonally and inter annually in response to changes in the heating and wind regimes. Seasonally, the winter circulation is characterized by a prevalence of warmer Mediterranean inflow reinforced by southerly winds. In summer, there is a slightly stronger outflow of fresher and warmer Adriatic water (Cushman-Roisin, 2001).

Mainly during the winter, the Adriatic Sea region is under a continuous influence of passing mid-latitude meteorological perturbations and of the wind systems associated with them. The two main wind systems are the Bora and the Scirocco, although Adriatic seawater circulation is also influenced by the local wind Maestral, the northwesterly wind typical of the summer season in the Adriatic. The Bora is a dry and cold wind blowing in an offshore direction from the eastern coast. The Scirocco blows from the southeast (i.e. along the longitudinal axis of the basin) bringing rather humid and relatively warm air into the region. In particular, the Bora produces appreciable buoyancy fluxes through evaporative and sensible heat loss, induces both wind-driven and thermohaline circulation, and, what is most important, is responsible for deep water formation processes (Leksikografski institut, 1979; Cushman-Roisin, 2001). Namely, the Adriatic Sea being exposed to very low winter temperatures and violent episodes of the Bora wind, has been identified as one of the regions of the world oceans where deep water formation processes take place. Meteorological conditions favourable to dense water formation cause rapid mixing of surface waters with deeper water layers. This dense water, called Adriatic Deep Water (ADW) spreads through the Strait of Otranto, being an important component of the Eastern Mediterranean Deep Water (EMDW) and contributing to their ventilation (Ovchinnikov et al., 1985).

*1.3 Radioactivity in the marine environment*

Radionuclides are of interest to marine scientists for two primary reasons: a) as potential contaminants of the ocean biosphere and b) as radioactive tracers for studies of water masses, sediment movements and various other parameters. The dominant route for the introduction of artificial radionuclides into the environment, until the nuclear accident in Chernobyl on 26 April 1986, has been the radioactive fallout resulting from the atmospheric nuclear weapon tests. Atmospheric nuclear explosions have been conducted since 1945 and were specially intensive in 1960s, i.e., before a nuclear moratorium became effective. However, similar, but smaller tests were performed by the Chinese and French also in the 1970s and afterwards. Therefore, activity concentrations of fission products in most of the environmental samples could be expected to be in good correlation with fallout activity (i.e., surface deposit in $Bqm^{-2}$). Among the man-made, i.e., anthropogenic radionuclides present in global fallout, $^{137}Cs$ and $^{90}$Sr have been regarded as the fission products of a major potential hazard to living beings due to the unique combination of their relatively long half-lives, and their chemical and metabolic properties resembling those of the potassium and calcium respectively.

The addition of artificial radioactive material to the marine environment inevitably results not only in increased radiation exposure to a marine biota, but also to some added radiation exposure to people who use the sea and its products. Consequently, concern for safety and radiation protection has worldwide stimulated much basic research dealing with radioactivity in the marine environment. Such investigations take significant part in an extended and still ongoing monitoring programme of radioactive contamination of human environment in Croatia as well (Popovic, 1963 - 1978; Bauman et al., 1979 - 1992; Kovac et al., 1993 - 1998; Marovic et al., 1999 - 2002).

*1.4 Tracer studies in the Adriatic Sea*

The investigations of radionuclides in the marine environment include studies of man-made ones (fallout radionuclides as well as radionuclides released from various nuclear facilities) that are already present in water, or by releasing specific radionuclides into the water in tracer experiments. It should be noted that for the purpose of a tracer experiments can be used also some non-radioactive elements or substances. In the Eastern Mediterranean Roether et al. have been conducting measurements of transient tracers (such as CFCs, helium and tritium) for more than last two decades (Roether and Schlitzer, 1991; Roether et al., 1994.) Initially they have been used to study deep water formation rates in the Adriatic and to identify the spreading pathways to the Mediterranean as well as to study their representation in models. At the end of the 1980s the thermohaline circulation of the Eastern Mediterranean changed abruptly causing the Aegean sea to replace the Adriatic sea as a much stronger source of deep water formation. This major event has been named Eastern Mediterranean Transient (EMT) and was attributed to important meteorological anomalies in the region as well as to changes in circulation patterns. The subsequent tracer surveys have been very valuable in documenting the ongoing-changes. Although the initial perturbation of the thermohaline circulation was rather short-lived, the effects are apparently long-lasting on time scales of decades. Water mass variability induced by the EMT was originally largest in the deep water but is now present in all water masses. Most recent tracer data from 2002 demonstrate that at present the strongest changes are to be found in the intermediate waters.

Regarding the long-term behaviour of fallout radionuclides in the Eastern Mediterranean, Franic and Bauman studied radioactive contamination of the Adriatic sea by $^{90}$Sr and $^{137}$Cs. Papucci et al. and Delfanti et al. analysed existing and new data on $^{137}$Cs distribution in the Eastern Mediterranean (Papucci and Delfanti, 1999; Delfanti, 2003). In recent study Sanchez-Cabeza et al. (2002) performed long-term box modelling of $^{137}$Cs in the Mediterranean sea. The main sources of the $^{137}$Cs to the Mediterranean area are fallout from past nuclear weapon testing and the Chernobyl accident in 1986. From 1963 to 1986, the $^{137}$Cs activity concentrations in surface water had regularly decreased, reflecting the decrease of the atmospheric input and vertical transport processes. The $^{137}$Cs vertical profiles were characterised by decreasing concentrations from surface to bottom. In the Eastern Mediterranean, the only vertical profiles before the EMT were obtained by Fukai et al. in mid seventies (Fukai et al, 1980). $^{137}$Cs surface concentrations ranged between 4.2 and 5.5 Bq m$^{-3}$ over the entire basin (Fukai et al., 1980) including the Adriatic sea (Franic and Bauman, 1993). The profiles showed subsurface maxima in the upper 400 m and an exponential decrease towards the bottom. It is reasonable to assume that the spatial distribution of $^{137}$Cs was similar to that of a tritium in 1978 (Roether and Schlitzer, 1991.), which showed concentrations below 1500 m decreasing from the Western Ionian Sea towards the Levantine basin. These distributions reflected the circulation in the Eastern Mediterranean, with tracer-rich dense waters of Adriatic origin flowing first into the bottom of the Western Ionian Sea and then spreading eastward. In 1986 the fallout from the Chernobyl accident produced a sharp increase in $^{137}$Cs concentration at the surface of the Eastern and Northern Mediterranean basins. However, the pre-Chernobyl levels were reached again in 1990.

In 1995, the vertical profiles of $^{137}$Cs at the two sides of Crete showed that the new deep water of Aegean origin was marked by relatively high concentrations of $^{137}$Cs ranging from 2 to 2.5 Bq m$^{-3}$ (Papucci and Delfanti, 1999). The vertical profiles in the Ionian and Levantine Seas showed surface concentrations around 3 Bq m$^{-3}$. However, in the bottom layer, they significantly differed from the old profiles. In the Ionian Sea, minimum concentrations (1-1.5 Bq m$^{-3}$) were found in the depth interval 750-1500 m, followed by an increase up to 2.5 Bq m$^{-3}$ from 2000 m to the bottom. These results indicate that the deep layer was still characterised by the presence of EMDW, and the increase in $^{137}$Cs concentration was due to its continuous transport from surface to bottom through convection processes in the Adriatic Sea.

It should be noted that the long-term data on $^{90}$Sr activity concentrations in sea waters worldwide are very scarce or non existent, which can mainly be attributed to long and tedious sample preparation and complex measurement procedure. In addition, due to relatively constant $^{137}$Cs:$^{90}$Sr activity ratios in environmental samples prior to the Chernobyl accident, generally ranging between values 1 and 3, for pure monitoring purposes it was straightforward to calculate $^{90}$Sr activity concentrations from the $^{137}$Cs obtained by gammaspectrometric measurements. Since both of these nuclides have inert gaseous precursors in their fission chains and similar radiological half-lives, substantial fractionation from the time of their creation by atmospheric nuclear weapon tests is considered unlikely. The expected value of this ratio in global fallout, for the pre-Chernobyl period, computed by Harley et al. (1965), based upon measured fission product yields was about 1.45. Thus, all of the fallout entering the sea was assumed to carry approximately this ratio of $^{137}$Cs to $^{90}$Sr. In the Adriatic sea, the mean value of $^{137}$Cs to $^{90}$Sr ratio in the period 1978 - 1985 was relatively constant, being 1.52 ± 0.40 (Franic and Bauman, 1993), which is in reasonable agreement with the values 1.5 - 1.6 determined for other seas (Volchok et al., 1971; Kupferman et al., 1979). As the consequence of the nuclear accident at Chernobyl, this ratio has been notably altered. Namely, since refractory components of the Chernobyl debris were deposited closer to the accident location than the more volatile constituents, due to the volatile nature of cesium and the refractory nature of strontium the $^{137}$Cs:$^{90}$Sr activity ratio in the Adriatic region significantly increased. Also, $^{137}$Cs deposition patterns around Europe were extremely nonuniform (EC, 2001). Consequently, although data on $^{137}$Cs activity concentrations in the sea waters worldwide are available, after the Chernobyl accident it was not possible from these data to estimate $^{90}$Sr activities.

In this paper are presented the results of long-term systematic measurements of $^{90}$Sr activity concentrations in the Adriatic sea water and fallout, which were then used to estimate its mean residence time in the Adriatic sea water, reflecting also the turnover time of sea water in Adriatic itself. Namely, in physical oceanography $^{90}$Sr can be used as radioactive tracer to study water mass transport, due to apparent constancy of stable strontium in the sea at relatively high concentrations of about 8 mg L$^{-1}$. However as the levels of $^{90}$Sr activity concentrations in the sea water decreased, studies involving that radionuclide became much scarcer since they involve long, tedious and costly radiochemical procedures.

*1.5. Literature data on water exchange between the Adriatic and the Ionian seas and turnover time of the Adriatic sea water*

The water exchange through the Strait of Otranto between the Adriatic and the Ionian sea has been the subject of a series of experimental investigations and more recently also of some numerical studies, which is extensively presented in Cushman-Roisin et al. (2001). From the data on water fluxes through strait can be easily calculated the turnover time of the Adriatic sea water by calculating the annual water mass flowing through the strait and dividing it by the total volume of the Adriatic sea.

Earlier results on the volume transport of sea water through the Strait of Otranto that had been obtained from very limited and sporadic current measurement, lead to the estimate of turnover time of approximately 5 years (Leksikografski institut, 1979). However, as the measured values of volume transport of sea water on certain spots had been in range of $2 \times 10^3$ to $5 \times 10^5$ m$^3$ s$^{-1}$ (Leksikografski institut, 1979), that called for further investigations. Estimates by Zore-Armanda and Pucher-Petrovic (1976) gave flux of $4.05 \times 10^5$ m$^3$ s$^{-1}$, leading to the turnover time of 2.7 years. It should be noted that this flux has been estimated for wintertime. Mosetti (1983) computed water transport through strait and obtained values ranging from $3 \times 10^5$ to $1 \times 10^6$ m$^3$ s$^{-1}$, which corresponds to the respective turnover times of 3.7 and 1.1 years, with the best estimate around $4 \times 10^5$ m$^3$ s$^{-1}$, which corresponds to the turnover time of 2.8 years. Summer flux that has been estimated by Orlic et al. (1992) to be $2.52 \times 10^5$ m$^3$ s$^{-1}$, corresponds to the turnover time of 4.4 years.

The flux estimates obtained from the direct current measurements data, and by numerical integration from the vertical distribution of the mean seasonal inflowing current component at the Otranto transect lead to annual mean volume transport of $1.09 \times 10^3$ to $5 \times 10^6$ m$^3$ s$^{-1}$, i.e. to turnover time of 1.0 years. This values seem more consistent with ADW formation rates as well as ADW outflow rates (Cushman-Roisin et al., 2001).

In recent study Sanchez-Cabeza et al. (2002) performed long-term box modelling of $^{137}$Cs in the Mediterranean sea. The prediction values of their numerical code, validated against literature data on $^{137}$Cs activity concentrations in the sea waters, were quite satisfactory in keeping with observations. The generic value for a total outflow from the Adriatic sea to the Ionian sea in that model was $3.8 \times 10^5$ m$^3$ s$^{-1}$, which leads to the turnover time of the Adriatic sea water of 2.9 years.

It should be noted that all above values for the water transport through the strait appear to be rather small, leading to overestimation of turnover time, considering that estimates of the average annual rate of Adriatic deep water formation are about $3 \times 10^5$ m$^3$ s$^{-1}$ (Roether and Schlitzer, 1991; Lascaratos 1993) and that ADW makes only one part of the total volume of the Adriatic waters exchanged through the Strait of Otranto.

Vetrano et al. (1999) estimated the annual mean volume transport during year 1995 to be $(1.11 \pm 0.44) \times 10^6$ m$^3$ s$^{-1}$. This has been done by numerical integration from the vertical distribution of the mean inflowing current component at Otranto transect. Therefore this results seem to be the most consistent with ADW formation rates. These water fluxes corresponds to the Adriatic sea water turnover time of 0.7 to 1.7 years.

In conclusion, literature data for Adriatic sea water turnover time range from 0.7 to 5 years, although recent estimates obtained from direct current measurements in the Strait of Otranto are on the order of 1 year.

**2. Material and methods**

*2.1. Sampling and radioactivity measurements*

$^{90}$Sr has been analysed in the sea water samples and fallout in the Adriatic since 1963. Sea-water samples, 150 - 200 L each, were collected twice a year (in May and

October, if feasible) 3 km from the shore, at a depth of 0.5 m, at four sampling locations of the Adriatic Sea: towns of Rovinj and Rijeka in North Adriatic, town of Split in Mid-Adriatic and town of Dubrovnik in South Adriatic. Reference methods for collecting procedure and handling of the sea water samples were taken from International Atomic Energy Agency reference book (IAEA, 1970). Fallout samples were collected monthly in the town of Zadar (Mid-Adriatic). The funnels which were used for fallout collection had a 1 $m^2$ area. Precipitation height was measured by Hellman pluviometer. Sampling sites and their coordinates are given in Fig. 1.

For the determination of strontium in fallout and sea water were used radiochemical methods (U. S. Department of Energy, 1957 - 1997; Bauman, 1974).

The radioactivity of $^{90}$Sr was determined by beta-counting its decay product, $^{90}$Y, in a low-background, anti-coincidence, shielded Geiger-Müller counter. Counting time depended on $^{90}$Sr activity concentration in samples, but was never less than 60,000 s, typically being 80,000 s.

Quality assurance and intercalibration of radioactivity measurements were performed through participation in the IAEA and World Health Organization (WHO) international quality control programmes.

It should be noted that the sampling locations were chosen not for the purpose of this investigation, i.e. studies of $^{90}$Sr mean residence time in the Adriatic and water mass transport, but as a part of an extended monitoring programme of radioactive contamination of Croatian environment.

*2.2. $^{90}$Sr activity concentrations in sea water and fallout*

Measured $^{90}$Sr activity concentrations in the Adriatic sea were found to be approximately equal on all observed locations, not differing from the rest of the Mediterranean sea (Franic and Bauman, 1993). It was estimated that approximately 85% of all man-made radioactive contamination in the Mediterranean comes from fallout (UNEP, 1991). Consequently, in the Adriatic, $^{90}$Sr activity concentrations are in good correlation with the fallout activity, the coefficient of correlation being 0.72 (Franic and Bauman, 1993).

As indicated on Figure 4, an exponential decline of $^{90}$Sr activity concentrations in sea water followed the nuclear moratorium in 1960s. $^{90}$Sr activity concentrations in the Adriatic sea water dropped from $14.8 \pm 2.4$ Bq m$^{-3}$ in 1963 to $2.0 \pm 0.3$ Bq m$^{-3}$ in 2003. Over the entire study period of 41 years, the maximum and minimum activities, 17.5 and 0.9 Bq m$^{-3}$, were detected in Split in 1963 and in Dubrovnik in 1983. In the same period, the total annual land surface deposit in Zadar fell by three orders of magnitude, from 713.3 Bq m$^{-2}$ in 1963 to 0.4 Bq m$^{-2}$ in 2003. The integrated delivery of $^{90}$Sr through atmospheric fallout for the period 1963 - 2003 was 2.91 kBq m$^{-2}$. If this value is considered with respect to the Adriatic sea surface area of $1.386 \times 10^{11}$ m$^2$ (Leksikografski institut, 1979) an estimate of $4.04 \times 10^{14}$ Bq is obtained as the total fallout delivery of $^{90}$Sr into the Adriatic sea for that period.

It should be noted that the Chernobyl accident did not cause any significant increase of $^{90}$Sr activity concentration in sea water, as well as in most of the environmental samples in Croatia, the exception being water samples from cisterns collecting rainwater from roofs etc. (Franic et al., 1999). Unlike the atmospheric testing of nuclear weapons, the radionuclides that originated from the Chernobyl accident were not released directly to the upper atmosphere. As the result of a release mechanism (prolonged burning of a graphite moderator from a damaged nuclear reactor) and prevailing meteorological conditions at that time, the less volatile components of Chernobyl debris (e.g., $^{90}$Sr) were deposited closer to the accident location than the more volatile constituents (e.g., $^{137}$Cs) (Aarkrog, 1988). Thus, $^{90}$Sr was only in minor quantities subjected to global dispersion processes as it was deposited to the surface of Earth within a period of few weeks after the accident. In addition, changing meteorological conditions with winds blowing from different directions at various altitudes and prolonged release from a damaged reactor resulted in very complex dispersion patterns over Europe. Consequently, the Adriatic region, except from the very northern and very southern part, was initially unaffected by the plumes of contaminated air (UNSCEAR, 1988; UNEP, 1991). Also, the late spring and early summer of 1986 in Croatia were rather dry, leading to relatively low direct radioactive contamination, which was especially true for the Adriatic region (Bauman et al., 1979-1992). Therefore, the average $^{90}$Sr activity concentration in the Adriatic sea water in 1986 was $2.1 \pm 1.5$ Bq m$^{-3}$. The minor increase was detected only in the sea-water sample collected in Dubrovnik in May 1986, with $^{90}$Sr activity concentration of 4.7 ± 0.4 Bq m$^{-3}$. However, that sampling coincided with a heavy rain in Dubrovnik region. For comparison, the $^{90}$Sr activity concentrations reported for the Black and Aegean seas in 1987 were significantly greater, 17.0 - 77.7 Bq m$^{-3}$ and 3.7 - 13.3 Bq m$^{-3}$ respectively (Polikarpov et al., 1991).

## *2.3. Model of the $^{90}$Sr circulation in the Adriatic sea*

To establish a simple mathematical model of the $^{90}$Sr circulation in the Adriatic sea water, the whole Adriatic was considered to be a single, well-mixed water reservoir, although it is probably not entirely true for the deep waters from the pits. However, the volume of these waters is small compared to the volume of the entire Adriatic. The $^{90}$Sr coming into the Adriatic sea by fallout and runoff due to the rapid mixing of surface waters with the intermediate water layer, together with the surface waters, sinks to deeper water layers and eventually, due to the general circulation pattern, leaves the Adriatic. Also, the assumption was made that the levels of $^{90}$Sr activity concentrations in Adriatic and Ionian sea waters were approximately the same.

Then, the rate of change of $^{90}$Sr activity concentrations in the Adriatic sea can be described by the simple mathematical model:

$$dA_{AS}(t) / dt = - k_{eff} A_{AS}(t) + I(t) \qquad (1)$$

where:

$A_{AS}(t)$ is the total, time-dependant $^{90}$Sr activity (Bq) in the Adriatic sea calculated from measured $^{90}$Sr activity concentrations in the sea water,

$1/k_{eff}$ the observed (effective) mean residence time of $^{90}$Sr (y) in a reservoir and

$I(t)$ the total annual $^{90}$Sr input to the Adriatic sea (Bq y$^{-1}$).

However, the observed constant which describes the rate of decreasing of $^{90}$Sr activity concentration in the sea water has to be corrected for the $^{90}$Sr radioactive decay. Therefore:

$$k_{eff} = - k_S + \lambda \qquad (2)$$

where:

λ      is the decay constant for $^{90}$Sr, i.e., 0.0238 y$^{-1}$ (ICRP, 1988) and

$1/k_S$   the mean residence time of $^{90}$Sr in the Adriatic sea (y).

The total, time-dependant $^{90}$Sr activity in the Adriatic sea can be calculated by multiplying the observed activity concentration data in the sea water by the volume of the reservoir:

$$A_{AS}(t) = A_{ASo}(t)\ V_{AS} \qquad (3)$$

where:

$A_{AS}(t)$  is the total, time-dependant $^{90}$Sr activity (Bq) in the Adriatic sea,

$A_{ASo}(t)$ observed (measured) time-dependant $^{90}$Sr activity concentrations (Bq m$^{-3}$) in the Adriatic sea for respective years and

$V_{AS}$    the volume of the Adriatic sea ($3.5 \times 10^{13}$ m$^3$).

The input of $^{90}$Sr to the Adriatic sea consists of three main components: fallout, runoff and influx of water into Adriatic through the Strait of Otranto. Input by fallout was estimated by multiplying the observed fallout data (Bqm$^{-2}$) by the area of the Adriatic sea ($1.386 \times 10^{11}$ m$^2$). The assumption was made that fallout data obtained from the measurements in the city of Zadar are fair representations for the whole Adriatic area, which might not be necessarily true. The time dependent $^{90}$Sr input I(t) in the Adriatic sea by fallout can therefore be modelled as:

$$I_f(t) = I_f(0) \exp(-k_f t) = S\ D_f(0) \exp(-k_f t) \qquad (4)$$

where:

$I_f(0)$   is the initial input of $^{90}$Sr in the Adriatic sea by fallout (Bq),

$k_f$     a constant describing the rate of annual decrease of $^{90}$Sr activity concentration in fallout (y$^{-1}$),

S      the area of the Adriatic sea ($1.386 \times 10^{11}$ m$^2$) and

$D_f(0)$  the initial $^{90}$Sr annual surface deposit by fallout per unit area (Bq m$^{-2}$).

By fitting the fallout experimental data for the 1963 - 2003 period (Fig. 3), for $k_f$ is obtained 0.388 y$^{-1}$ and for initial value, i.e., D(0), was obtained 112.6 TBq, which corresponds to the surface deposit of 812 Bq m$^{-2}$ for the year 1963.

**Figure 3 about here**

As can be seen from the Fig. 3, the rate of decrease of $^{90}$Sr activity concentration in fallout in 1960s is very fast, slowing ever since. From the exponential curve, the mean residence time of $^{90}$Sr in fallout, $1/k_f$, was calculated to be about 3 years.

To estimate runoff contribution to the input, it is reasonably to assume that the $^{90}$Sr activity concentrations in runoff depend upon activity concentrations in fallout, since correlation has been found between fallout and sea water activities, as well as between fallout and activities of fresh water (Franic and Bauman, 1993; Bauman et al., 1979 - 1992). The long-term average runoff rate along the Adriatic coast is 5500 - 5700 m$^3$ s$^{-1}$ (Cushman-Roisin et al., 2001), the biggest contributor of fresh water being the Po river, with annual mean runoff of about 1,700 m$^3$ s$^{-1}$. In addition, a land runoff, which is not collected into rivers, was estimated to be 1100 m$^3$ s$^{-1}$ (Cushman-Roisin et al., 2001). The total runoff of 6800 m$^3$ s$^{-1}$ adds annually about 215 km$^3$ of fresh water to the Adriatic. Assuming the $^{90}$Sr activity concentrations in runoff to be similar to those in sea water, runoff would on annual basis add less than 3% to the $^{90}$Sr input to the Adriatic sea. Consequently, the initial value for $^{90}$Sr input by fallout into the Adriatic, adjusted for runoff, is 116 Tbq.

Finally, the amount of $^{90}$Sr entering the Adriatic sea by influx of sea water through the Strait of Otranto is proportional to the $^{90}$Sr activity concentrations (Bq m$^{-3}$) in the Ionian sea and the volume of water entering the Adriatic. Assuming that the mean residence time of $^{90}$Sr in the Adriatic sea also reflects the turnover time of the Adriatic sea water, the annual activity of $^{90}$Sr entering the Adriatic through the Strait of Otranto would be proportional to the volume of sea water passing through the Strait of Otranto into the Adriatic, which is equal to the volume of the Adriatic sea divided by the mean residence time, i.e. $V_{AS} / T_M = V_{AS} \times k_S$. In the Adriatic sea this activity exponentially decreases with the constant $k_{eff}$ described by the equation (2). Therefore, $^{90}$Sr activity entering the Adriatic through Strait of Otranto can be described by equation:

$$I_{IS}(t) = V_{AS}\, k_S\, A_{ISo}(0)\, \exp(-k_{eff}\, t) \qquad (5)$$

where:

$I_{IS}(t)$ is the annual input of $^{90}$Sr to the Adriatic sea from Ionian sea (Bq) and

$A_{ISo}$ the initial observed activity concentration of $^{90}$Sr in the Ionian sea (Bq m$^{-3}$).

After combining equation (1) with (2), (3), (4) and (5) and solving for $A_{AS}(t)$ with initial conditions $I_f(0) = I(\text{year 1963})$ and $A_{ASo}(0) = A_{ASo}(\text{year 1963})$, the following solution is obtained:

$$A_{AS}(t) = I_f(0) / (k_S + \lambda - k_f) \{ \exp(-k_f\, t) - \exp[-(k_S + \lambda)\, t] \} + \qquad (6)$$

$$+ V_{AS}\, [A_{ASo}(0) + A_{ISo}(0)\, k_S\, t]\, \exp[-(k_S + \lambda)\, t]$$

Equation (6) is then fitted to the $^{90}$Sr activity data in the Adriatic sea water obtained from the equation (3). As $^{90}$Sr data for the Ionian sea are unavailable, it was assumed that the initial $^{90}$Sr activity concentration in the Ionian sea was approximately equal to its activity concentration in the Adriatic.

## 3. Results and discussion

### 3.1. The mean residence time of $^{90}$Sr in the sea water

By fitting the experimental data to a theoretically predicted curve (6), the unknown parameter $k_S$ was calculated to be 0.307 y$^{-1}$. Although the model (6) is simplified representation of the real situation, the fit is reasonably good, the coefficient of correlation between experimental and predicted data being 0.74. However, as shown on Fig. 4, the model under represents the real data on the far end (i.e. from early 1980s) since by that time the activity concentrations both in fallout and sea water became very low, essentially reaching equilibrium.

**Figure 4 about here**

The mean residence time of $^{90}$Sr in the Adriatic, being the reciprocal value of $k_S$

was estimated to be approximately $T_M$ = 3.3 years. As $^{90}$Sr is an effective radioactive tracer of water mass transport, this value also reflects the turnover time of the Adriatic sea water. However, this value could be regarded as an upper limit averaged over the period of 40 years.

The value of 3.3 years is approximately 4 times smaller compared to the observed residence time of $^{90}$Sr in the Adriatic sea, which was calculated, by fitting the observed $^{90}$Sr activity concentrations in sea water to exponential curve, to be 12.4 years (Franic and Bauman 1993).

*3.2. Error estimation and sensitivity analysis*

In order to obtain the standard deviation of $T_M$, Monte Carlo simulations were performed. To be on a conservative side, as well as to simplify calculations, the uniform distribution has been assumed over the A ± s value of $^{90}$Sr sea water activity concentrations for respective years, although normal would be more realistic. For each year the random value was generated over the interval [A - s, A + s] and then from such set of data was estimated the $1/k_S$ value by fitting to the equation (6). The process has been repeated 100 times and 100 values for $1/k_S$ were obtained. The mean value and standard deviation for $T_M = 1/k_S$ were calculated to be 3.3 ± 0.4 years.

In order to estimate which parameter from the equation (6) mostly affects the final result, was performed sensitivity analysis. Sensitivity analysis involves perturbing each parameter of model by a small amount, while leaving all other parameters at nominal (preselected) values and quantifying the relative effect on the model prediction. Usually it is performed by increasing or decreasing each parameter over its entire expected range by a fixed percentage of the nominal value. In the case of model (6) the parameters that affect the final result are strontium input into the sea water by fallout and runoff, input from the Ionian sea (which itself depends upon $^{90}$Sr activity concentration in Ionian sea and water flux through the Strait of Otranto) and $^{90}$Sr activity concentrations in the Adriatic sea. As previously noted, the sampling locations for the long-term investigations of $^{90}$Sr in the Adriatic were not chosen for the purpose to estimate of the total content of $^{90}$Sr in the Adriatic sea needed for studying mean residence time of $^{90}$Sr and water mass transport, but as a part of an extended monitoring programme of radioactive contamination of Croatian environment. In addition, $^{90}$Sr data for the Ionian sea are unavailable. The range over which were varied each of critical parameters in the model, i.e., $A_{AS}$, $A_{IS}$ and $I_f$, was arbitrarily chosen to be ±25% around the nominal value.

Increasing the $^{90}$Sr input by fallout 25%, equation (6) yields for the mean residence time of $^{90}$Sr in sea water value of 3.0 years. On the other side, by decreasing fallout input 25%, the mean residence time of 3.6 years is obtained. Implementation of similar procedure for the input of $^{90}$Sr through the water mass transport form the Ionian sea, yields the values of 2.9 and 3.7 years for a volume increase of 25% and volume decrease of 25% respectively. Finally, increase and decrease of $^{90}$Sr total activity in the Adriatic sea lead to respective turnover times of 3.5 and 2.9 years.

On Fig. 5 is shown how the mean residence time of $^{90}$Sr in the sea water depends upon variation of those three parameters over their default values.

**Figure 5 about here**

As seen from fig. 5, ± 25% uncertainty in estimation of the Adriatic sea water activity causes approximately -10% and +10% change in $^{90}$Sr mean residence time. This uncertainty arises from the fact that only four sampling locations were used for estimation of the $^{90}$Sr activity for the total volume of the Adriatic sea. On the other hand, the larger input of $^{90}$Sr either by fallout or by water influx from Ionian sea leads to smaller value for mean residence time.

As a consequence of direct proportionality between strontium input into the Adriatic sea and its mean residence time in the sea water, it can be argued that 3.3 is the upper limit of the Adriatic sea water turnover time. Namely, resuspension from the sediments could affect $^{90}$Sr activity concentrations, acting as additional input, especially in the northern, relatively shallow part of Adriatic.

The upper limit of 3.3 ± 0.4 years of the Adriatic sea water turnover time, estimated from long term observations of $^{90}$Sr as radiotracer, is in reasonable agreement with the overall literature data discussed in section 1.5.

## 4. Conclusions

Over the period of 41 years, $^{90}$Sr activity concentrations in the Adriatic sea water dropped from $14.8 \pm 2.4$ Bq m$^{-3}$ in 1963 to $2.0 \pm 0.3$ Bq m$^{-3}$ in 2003. In the same period, the total annual land surface deposit in Adriatic fell by three orders of magnitude, from 713.3 Bq m$^{-2}$ to 0.4 Bq m$^{-2}$. The Chernobyl accident did not cause any significant increase of in $^{90}$Sr activity concentration in sea water, as well as in most of the environmental samples in Croatia.

Using relatively simple mathematical model describing the rate of change of $^{90}$Sr activity concentrations in the Adriatic sea water, the mean residence time of $^{90}$Sr in the sea water was estimated to be $3.3 \pm 0.4$ years. However, turnover time of 3.3 years could be regarded as an upper limit averaged over the period of 40 years. As this value reflects the turnover time of the Adriatic sea water, for the spontaneous cleanup of well-mixed pollutant in Adriatic, theoretically it would take up to 3.3 years.

This value is comparable to the literature data for the values of the Adriatic sea water turnover that range from 0.7 - 5 years, obtained by studying water flows of the Adriatic sea water through the Strait of Otranto.

As the $^{90}$Sr activity concentrations in fallout and sea water are approaching background values, further improvements in any model that would use radiostrontium as a radioactive tracer are not very realistic.

**Acknowledgement**

This work received financial support from the Ministry of Science and Technology of the Republic of Croatia under grant # 00220204 (Environmental Radioactivity).

**Tables**

Table 1
Physical characteristics of the Adriatic Sea

|  | Adriatic Sea | North Adriatic | South Adriatic |
|---|---|---|---|
| Area | 138,600 km$^2$ | 78,750 km$^2$ | 59,850 km$^2$ |
| Volume | 35,000 km$^3$ | 7,000 km$^3$ | 28,000 km$^3$ |
| Volume of surface (mixed) layer | 4,200 km$^3$ | | |
| Volume of intermediate layer | 25,000 km$^3$ | | |
| Average depth | 160 m | | |
| Maximum depth | 1,220 km | | |

**Figures**

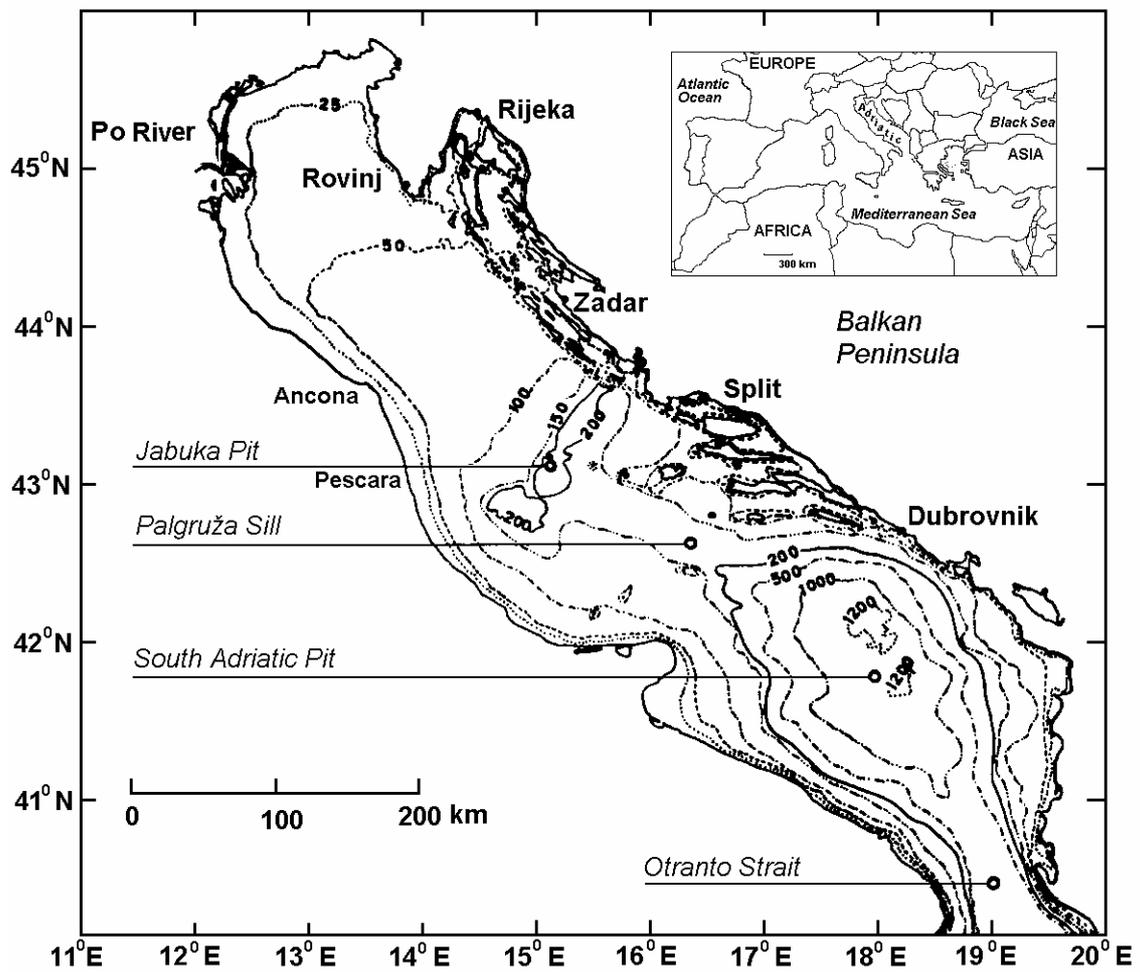

Fig. 1.
Map showing the position and topography of the Adriatic sea as well as sampling locations referred to in the text.

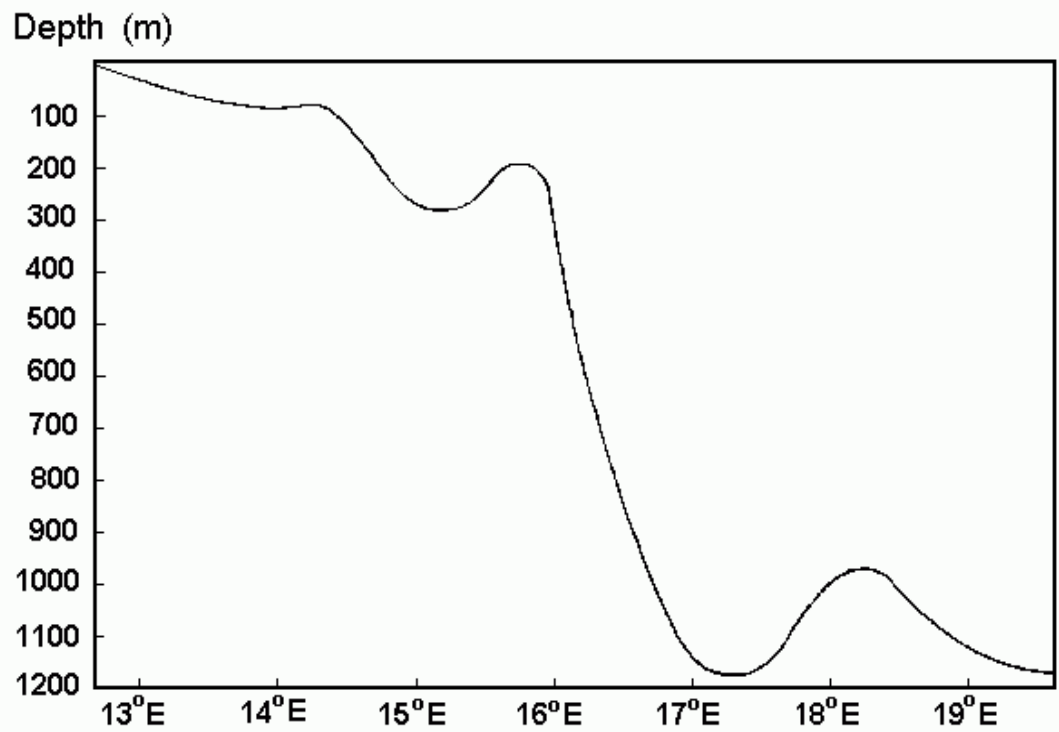

Fig. 2.
Bottom depth of the Adriatic sea along the NW - SE axis.

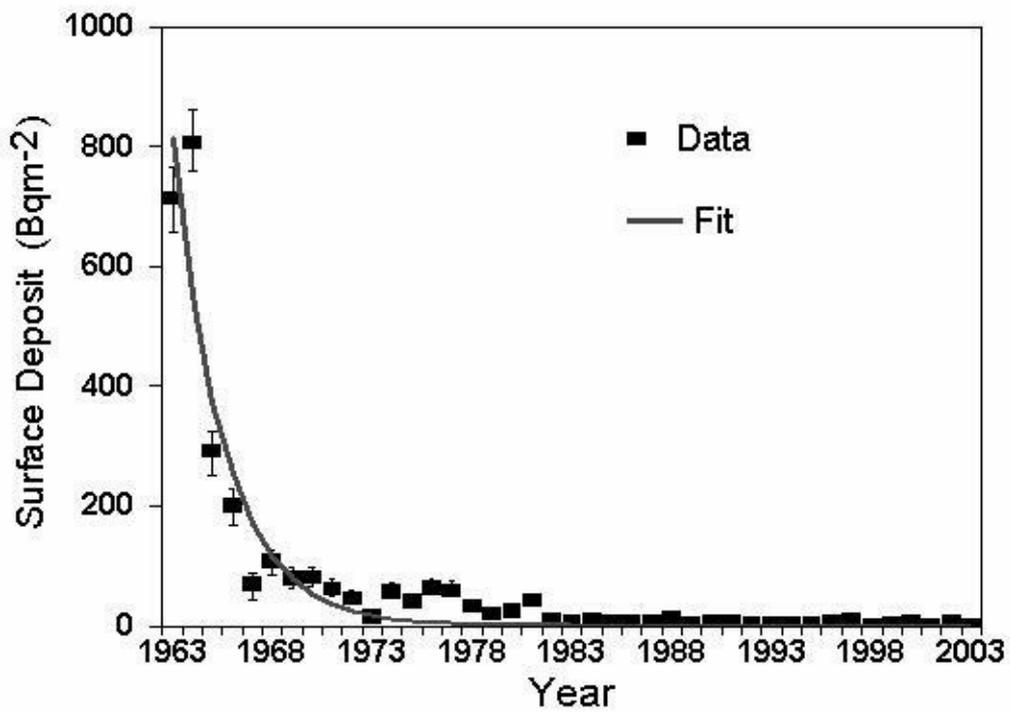

Fig. 3
$^{90}$Sr annual surface deposit calculated from $^{90}$Sr activity concentration in fallout collected in the city of Zadar from 1963-2001. Activity concentrations are reported as ± two sigma counting error.

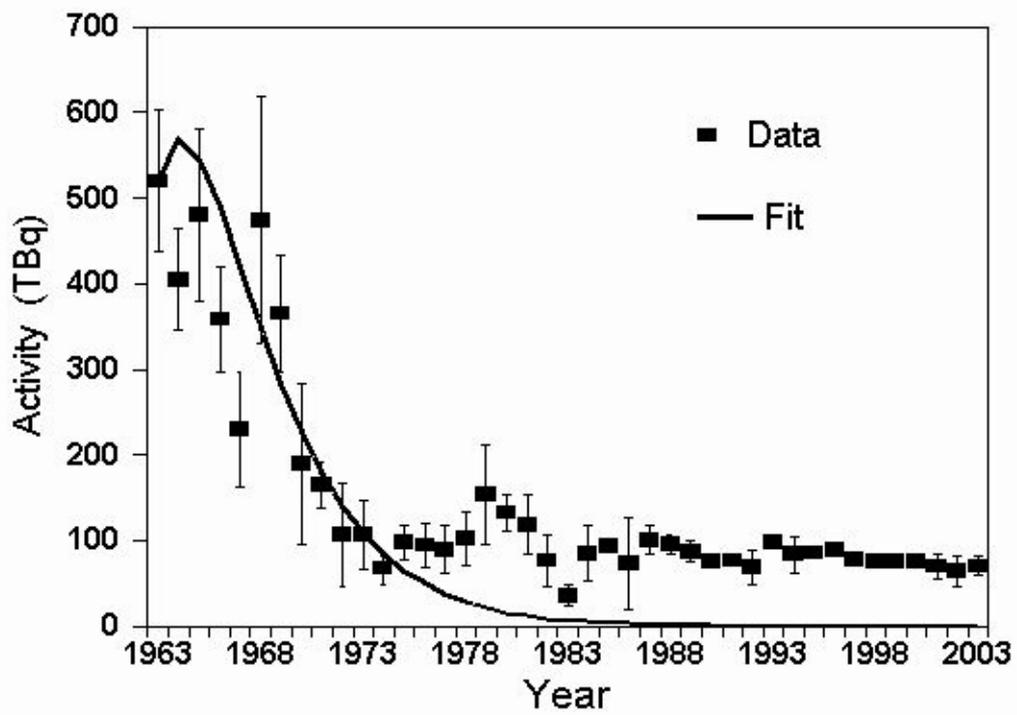

Fig. 4.
Mean value of $^{90}$Sr activity concentration of the Adriatic sea water on four sampling locations.

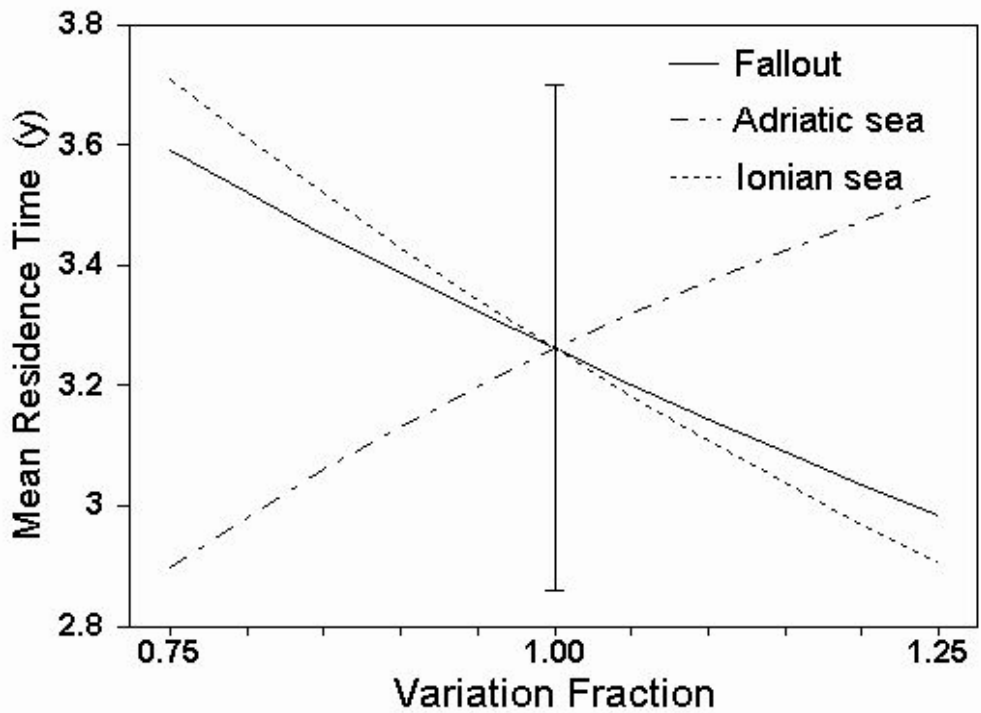

Fig. 5.
Sensitivity analysis of $^{90}$Sr mean residence time as a function of variation of strontium input in the Adriatic by fallout, input from the Ionian sea through Strait of Otranto and as a function of total activity in the Adriatic sea. Bar represents standard deviation of mean residence time obtained by Monte Carlo analysis.